# Selective Encryption using Segmentation Mask with Chaotic Henon Map for Multidimensional Medical Images

S Arut Prakash [1], Aditya Ganesh Kumar [2], Prabhu Shankar K. C. [3]
Lithicka Anandavel [4], Aditya Lakshmi Narayanan [5]

arutprakash2000@gmail.com,adty910@gmail.com,
prabushc@srmist.edu.in, lithicka@gmail.com, sai.9500@gmail.com

SRM Institute of Science and Technology, Kattankulathur, India

**Abstract.** A user centric design and resource optimisation should be at the center of any technology or innovation. The user centric perspective gives the developer the opportunity to develop with task-based optimisation. The user in the medical image field is a medical professional who analyses the medical images and gives their diagnosis results to the patient. This scheme having the medical professional user's perspective innovates in the area of Medical Image storage and security. The architecture is designed with three main segments namely - Segmentation, Storage and Retrieval. This architecture was designed owing to the fact that the number of retrieval operations done by medical professionals was toweringly higher when compared to the storage operations done for some handful number of times for a particular medical image. This gives room for our innovation to segment out the medically indispensable part of the medical image, encrypt it and store it. By encrypting the vital parts of the image using a strong encryption algorithm like chaotic Henon map we are able to keep the security intact. Now retrieving the medical image demands only the computationally less stressing decryption of the segmented region of interest. The decryption of the segmented region of interest results in the full recovery of the medical image which can be viewed on demand by the medical professionals for various diagnosis purposes. In this scheme we were able to achieve a retrieval speed improvement of around 47% when compared to a full image encryption of brain medical CT images.

## 1    Introduction

Medical imaging as stated by the United States FDA "refers to several different technologies that are used to view the human body in order to diagnose, monitor, or treat medical conditions. Each type of technology gives different information about the area of the body being studied or treated, related to possible disease, injury, or the effectiveness of medical treatment". These days, medical imaging is particularly centred around MRIs (Magnetic Resonance Imaging), CTs (Computer Tomography), PETs (Positron Emission Tomography), and Ultrasounds. Medical Imaging assists patients by recognizing diseases sooner and monitoring their treatment in a comprehensive manner.

Due to the wide-ranged observation made by doctors regarding the patients, a large amount of data is collected as medical images. This raises the concern of the security of the data. Patients' highly confidential information could be stolen and revealed by unauthorized parties, resulting in complications in patients' lives. It's possible that if the patient is a high-profile figure, even national security can be jeopardized. In light of these expanding dangers, we need to secure these medical images.

These security concerns have been recognized and various methods were planned and implemented but have had some concerns due to the aspects of medical images like redundancy and various conventionally implemented algorithms like AES, DES, RC4, etc. were better suited for textual encryption than image-based encryption.

Encryption can be accomplished in one of two methods. First, full encryption where the entire image is hidden. The second method is partial or selective encryption, which conceals only certain parts of an image. Since selective Encryption focuses on encrypting the most sensitive and significant parts of the data by adding noise to the parts requiring encryption, it is faster than full encryption. Images below show a comparison of source, full encryption, and selective encryption.

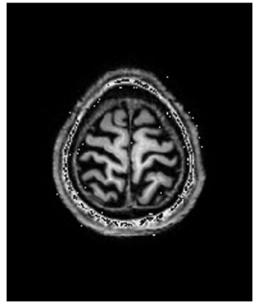 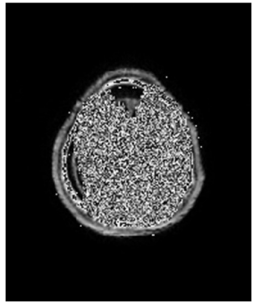 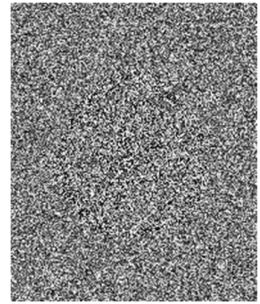

**Fig. 1.** Source Image    **Fig. 2.** Selective Encryption    **Fig. 3.** Full Encryption

This paper focuses on using Deep Learning-based Segmentation to extract Regions of Interest (ROI). After segmenting the ROI, it is encrypted using the chaotic map technique (Henon Map) which is generated using a key. Image is decrypted by using the same mask that was used for encryption and the same chaotic map is generated again using the same key. Our model with selective encryption is 47% faster than full encryption. In addition, we receive a lossless decrypted image.

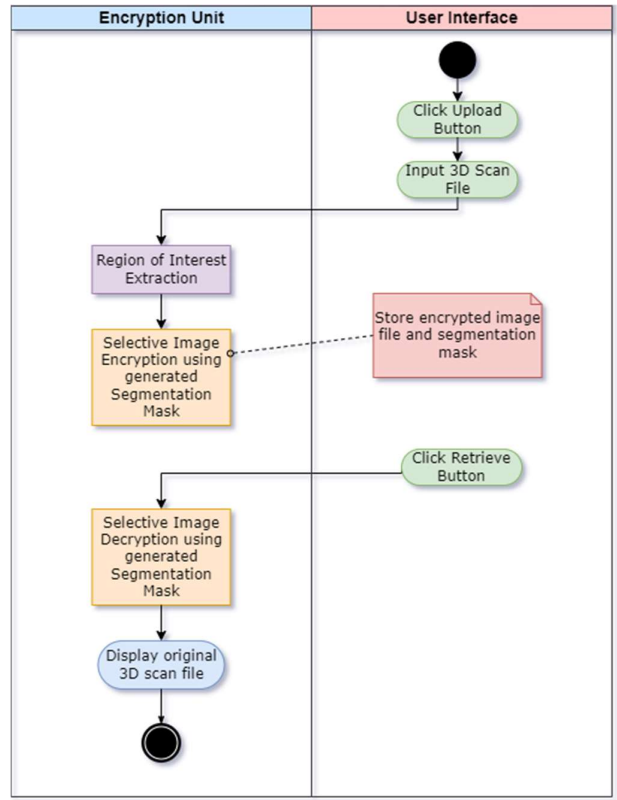

**Fig. 4.** Use case diagram

This framework is capable of operating on all types of Medical Images (NIfTI, DICOM) because the framework operates on the underlying image array. Retrieval of the medical image is faster as only selective regions (ROI) have to be decrypted.

## 2 Literature Survey

### 2.1 Image Segmentation

In the 2019 paper, Sihong Chen et al. explain the challenges faced trying to build a massive dataset for 3D Medical Image analysis and created 3DSeg-8 by combining multiple Medical Challenges datasets. This proved a wide array of modalities, target organs, and diseases to explore and work on. Using this, they built a diverse 3D network titled Med3D, which was used to co-train multi-domain 3DSeg-8 models to create a set of pre-trained models to generate the generic Medical 3D characteristics.

In 2018, Zongwei Zhou et al. showcased a novel UNet+ architecture for image segmentation. This involved using a deeply supervised encoder-decoder network architecture. The networks have sub-networks connected through a series of nested, dense skip-pathways. The paper suggests that when the encoder decoders are seman-tically similar the optimizer has an easier learning task. UNet++ with achieves IoU gains of 3.9 with U-Net and over 3.4 over wide U-Net.

## 2.2 Encryption

In their 2022 paper, Arshad et. al. used the concept of adjacency matrix from graph theory to generate non-linear S-boxes. Using the matrix with a Galois field, another S-box is generated. This combination is proved to be resilient to a variety of attacks like linear and differential attacks.

In 2021, Masood et. al. proposed three chaos based methods, Henon chaotic map, Brownian movement and Chen's chaotic system to encrypt images. The chaotic map generates random sequencing that is applied to the medical image. Each pixel in the image is distorted by first randomly shuffling them and finally bitwise XOR is applied. A symmetric key based cryptosystem is employed to encrypt as well as decrypt the medical image. This method requires less computational resources and offers faster processing. Deb et. al.'s paper from the same year proposes an encryption system using a non-linear filter function and shift registers. The medical image is randomized by a Logistic-Tent map and scrambled by Arnold transformation method. Next, the disordered image data is XORed with output sequences of specially designed PRNG to obtain a cipher image. Encryption decryption time requirements reveal the efficiency of the proposed system. Several performance measures are estimated to validate the resistance of the proposed scheme against statistical, differential, and a few common cryptanalytic attacks.

In 2020, S. Ibrahim et. al. developed a real-time encryption system. Since speed is of paramount importance here, Baker maps and Henon maps are selected as viable options. The novelty of the work is in the sandwiching of chaotic maps between S-box substitutions. This makes the framework resilient to chosen ciphertext, chosen plaintext and PRNG-reset attacks. Also, here the chaotic source is initialised using the AES algorithm. Through a variety of tests like differential attack test and cross-correlation analysis, Henon map is a better candidate for CT and X-Ray images, while Baker map is better for MRI images.

In 2018, the paper by Elhoseny discusses the use of traditional Elliptic Curve Cryptography for medical image encryption. But, to obtain the optimal key for the ECC, a novel algorithm based on the Grasshopper Optimisation (GO) and Particle Swarm Optimisation (PSO) is developed titled GO-PSO. The benefit of this method is that it is able to generate higher PSNR between source and encrypted image (which is desirable) in lesser number iterations than if either of the two methods were used in isolation. Also, the CPU time taken by this GO-PSO method is less than AES and RSA.

In their 2016 paper, Rehman et. al. used the Burger map for confusion operation of plaintext images. The output is then split into 64x64 blocks and then passed through a logistic map for substitution. This method is significantly faster than pre-existing approaches in terms of plaintext encryption. Also, the authors show that the key-space is large enough to warrant protection against brute force attacks, since two chaotic maps are being used here.

## 2.3 Selective Encryption

Osama A. Khashan et al. in the 2020 paper presented a lightweight solution to encrypting medical images that can be applied to real time applications. It is achieved by encrypting the edge maps of the image. The edge image blocks are obtained using an edge detection algorithm and a threshold value. Then, to construct a vast key space, a chaotic map is employed. The selected image blocks are encrypted using a one-time pad technique.This method is capable of offering a high level of security while reducing computation complexity.

The 2019 paper, Shahrokh Heidari et. al. presents a selective encryption for medical images based on quantum image processing. The suggested approach, which is based on BRQI representation, aims to encrypt a particular section of medical pictures called ROI (Region of Interest) by altering the order of its bitplanes in accordance with a key sequence. This method reduces the complexity significantly and the size of the medical image does not affect the approach. In the same year, the paper by L. Oteko Tresor and M. Sumbwanyambe proposes a selective image encryption algorithm based on 2D Discrete Wavelet Transform, Henon's Map and 4D Qi Hyper Chaos. The picture is decomposed using the DWT technique, the decomposed pixels are shuffled using the henon map, and the shuffled image is encrypted using bit streams produced by Qi hyperchaos using the XOR operation. The results of the experiments show that the suggested method has a wide key space, good security, and is resistant to many forms of assaults. When compared to various existing algorithms, the proposed algorithm outperforms them.

In their 2018 paper, Noura et. al. proposed three types of encryption algorithms to combat the problem of increased resources when processing latency in traditional encryption algorithms. These variants are full, middle-full and selective. This results in a highly secure encryption as the image is encrypted independently and uses a dynamic key. It is also lightweight since only one variable cipher primitives are applied to each image. By employing permutation of the sub-matrices of ROI lesser computational cost, lesser visual degradation and memory is obtained when compared to full encryption. This approach is best for real-time applications and systems with constrained devices.

Yang Ou et al. proposed two schemes of region-based encryption of medical pictures in a research published in 2017. The first method randomly flips the subset of the bits in the ROI's coefficients inside of several wavelet sub-bands. The second method involves encryption of a certain region's data in a code stream using AES. Both these methods can be employed to different medical images depending on the requirements. First scheme is optimal when the requirement is to have low impact on compression efficiency. When high security is desired, the second scheme can be used; it also does not modify the file size.

Muhammad Sajjad et al. in a 2016 publication proposed a system to outsource encryption to the cloud in order to tackle the problem of securely transferring a medical picture from a device with limited capabilities to remote patient monitoring centres. A visual saliency model detects the region that holds information in the medical image. An edge-directed data hiding method is then used to embed ROI into a host image. Due to the lower payload from selective encryption, a high-quality steganography image is produced. A high level chaotic encryption is employed at the cloud. The client retrieves the encrypted data and combines with the non region of interest image which can be sent to the medical professionals. This technique preserves image quality by using the aforementioned edge-directed method and decreases computational complexity by outsourcing the costly encryption procedure to the cloud.

### 2.4    Evaluation Metrics

The 2004 paper by Zhou Wang et. al. briefs the limitation on the existing error-sensitivity approach and proposes a quantitative measure that automatically assesses the quality of image with assumption that a full-reference image is available. They devised a structural similarity measure (SSIM) that analyses local patterns of pixel intensities that have been normalized for luminance and contrast. SSIM performs better than widely used reference quality metrics.

In 2002, the same authors had proposed a universal image quality index that can be applied across various image processing applications and provide comparison across different types of image distortions. Three factors are taken into consideration in this index, degree of correlation, mean luminance and contrast distortion. This mathematically defined method measures the features locally and combines them. This method outperforms widely used measurement methods like MSE and PSNR.

## 3    Preliminaries

In the preliminaries, the prerequisite knowledge required to understand the work is established. This section is divided into units elaborating on the datasets used, the underlying deep learning architecture used for region of interest extraction, the encryption method used and the evaluation metrics measured.

### 3.1    Datasets

**MRBrainS18.** The MRBrainS18 dataset is a dataset of the brain compiled by medical imaging experts for MICCAI 2018 Grand Challenge on MR Brain Segmentation.

The image data used here were obtained from a 3T scanner at the UMC Utrecht, Netherlands. 30 Subjects were taken for the data acquisition of the fully annotated scans having multi-sequence of the following -

- T1-weighted
- T1-weighted inversion recovery
- T2-FLAIR

Alzheimers, diabetes, dementia, white matter lesion in patients with age above 50, matched controls varying degrees of atrophy are some of the conditions of the 30 subjects. For each patient the following sequences were provided:

**Table 1.** MRI Sequences used in 3T scanner.

| Sequence | Details | TR (MS) | TE (MS) | TI (MS) |
|----------|---------|---------|---------|---------|
| T1 | 3D T1-weighted sequence | 7.9 | 4.5 | |
| T1-IR | Multi-slice T1-weighted inversion sequence | 4416.0 | 15 | 400.0 |
| T2-FLAIR | Multi-slice T2 flair sequence | 11000.0 | 125 | 2800.0 |

All the scans in the dataset are aligned and have a voxel size of 0.958mm × 0.958mm × 3.0mm. The N4ITK algorithm is used in correcting the bias fields in the scans. The T1-IR images also contain the artifacts at the bottom of the scans which is common in clinical scans.

The MRBrainS18 dataset contains 11 labelled categories. The labels were carried out by medical imaging experts using the manual reference standards. Each file in the dataset is of NIfTI format and volumetric in nature with dimensions approximately 140x230x195. The Labels include Background, Cortical gray matter, Basal ganglia, White matter, White matter lesions, Cerebrospinal fluid in the extracerebral space, Ventricles, Cerebellum, Brain stem, Infarction and Others.

Medically important segments could be any part of the brain in the scan. Securing the information/image of the brain with or without any medical condition is of utmost importance. For this study we use this dataset with 2 labels - Brain ROI and Background. The Brain ROI (Brain Region of Interest) is the region including all the labels other than background. Merging all the labels from 1-10 (Cortical gray matter, Basal ganglia, White matter, White matter lesions, Cerebrospinal fluid in the extracerebral space, Ventricles, Cerebellum, Brain stem, Infarction and Others) get the brain segment. So, in prediction 0 would account to background and 1 would account to the Brain ROI

**Table 2.** MRBrainS18 dataset label description.

| Label | Description |
|-------|-------------|
| 0 | Background |
| 1 | Cortical gray matter |
| 2 | Basal ganglia |
| 3 | White matter |
| 4 | White matter lesions |
| 5 | Cerebrospinal fluid in the extracerebral space |
| 6 | Ventricles |
| 7 | Cerebellum |
| 8 | Brain stem |
| 9 | Infarction |
| 10 | Other |

**3DSeg-8.** The 3DSeg-8 dataset was introduced in the Med3D paper by Chen et al. It has medical imaging of MRI and CT modalities compiled from 8 available 3D segmentation datasets. The dataset also includes various scan regions, organs and pathologies. Specifically, the dataset contains images of the brain, hippocampus, prostrate, liver, heart, pancreas, vessel and spleen. In total, the dataset contains 1638 cases. This huge variation in organ types paired with the varying modalities is intended to provide scale and task invariance to a network that is pre-trained on it.

### 3.2    3D ResNet

The ResNet model uses 3x3x3 convolutions with padding. The actual network used in the Med3D architecture consists of a backbone (encoder) along with convolutional layers (decoder) that follow this. The core backbone does not change for different situations and can be adapted to different tasks by just altering the convolutional layers. The efficiency of the Med3D model comes from the 3DSeg-8 dataset on which it was pre-trained upon. The reason Med-3D is able to succeed in transfer learning despite the wide variation in its pre-training data is due to 2 operations performed on the incoming data.

**Spatial Normalization.** This operation helps tackle voxel spacing distance (i.e.) the physical distance between two pixels in a 3D image. Also, for interpolation, median spacing is used. Since the physical distance varies with different domains, the Med3D employs a domain-specific median spacing.

For domain j, the median spacing for the $i^{th}$ image where ax denotes the x,y,z axes is defined by $sp^{ax}_{med, j}$

$$sp^{ax}_{med,j} = f_{med}(sp^{ax}_{0,n}, sp^{ax}_{1,n}, ..., sp^{ax}_{N_j,n})$$

Here, $f_{med}$ is the median operation performed on $N_j$ number of data in the jth domain.

Using $sp^{ax}_{med,\,j}$ we can proceed to calculate the new size of the image after spatial normalisation using the below formula.

$$s'_{ax,i,j} = \frac{s_{ax,i,j}}{sp^{ax}_{i,j}} \times sp^{ax}_{med,j}$$

**Intensity Normalization.** This operation is to tackle the variation in pixel values that are observed in medical images. After sorting the pixel values in the images, we remove pixel values outside the 0.5 to 99.5 percentile range. After this, the following operation is performed on all the pixel values $v_i$ in an entire volume

$$v'_i = \frac{v_i - v_m}{v_{sd}}$$

Here $v'_i$ is the final pixel value after intensity normalisation of the pixel i.

### 3.3 Henon Map

A dynamic system defined as "a deterministic mathematical model, where time can be either a continuous or a discrete variable." These models can either be studied as mathematical objects or used to describe a target system. A dynamical system is characterized as linear or nonlinear depending on the type of equations of motion used to describe the target system. If a system is linear, it will follow the principle of superposition, that is the equation will have a solution as $x_3(t) = ax_1(t) + bx_2(t)$, where a and b are constants. If a system is nonlinear it fails the principle of superposition, that is $x_3(t) = ax_1(t) + bx_2(t)$ cannot be shown as a solution, where a and b are constants. In addition, the nonlinear system does not change proportionately to a change in a variable. The dynamical systems of interest in chaos studies are nonlinear.

Chaos theory focuses on the behaviour of a dynamic system. It is the study focused on underlying patterns and random behaviours in a system governed by deterministic laws and highly sensitive to initial conditions in dynamic systems and has a dense set of periodic points. One popular example of chaos theory is the butterfly effect discovered by Meteorologist Edward Lorenz. He observed that even a basic heat convection model had inherent unpredictability and coined this phenomenon as the butterfly effect, suggesting that this phenomenon has the potential to cause a change in the weather like a hurricane or tornado with the mere flap of butterfly wings.

Chaotic maps are mathematical functions that produce random sequences. They can be parameterized by a discrete or continuous time parameter. Due to their diverse qualities such as ergodicity, mixing property, and sensitivity to beginning circumstances, they play an essential role in several fields. Some of the common examples of chaotic maps are Lorenz system, Arnold cat map and Henon map.

Henon map is one of the iterated dynamical systems that exhibit chaotic behaviour. The henon map follows a pair of two dimensional differential equations.

$$\begin{cases} x_{n+1} = 1 - ax_n^2 + y_n \\ \quad\quad y_{n+1} = bx_n \end{cases}$$

The map takes $x_n$ and $y_n$ as inputs and maps them to new points $x_{n+1}$ and $y_{n+1}$. The constants a and b are bifurcation parameters. The values 1.4 and 0.3 for a and b respectively will produce a chaotic map. Substituting different values might not produce a chaotic system.

### 3.4    Metrics

**Peak Signal to Noise Ratio (PSNR).** It is used to calculate the ratio between the maximum strength of the signal to that of data loss. It is used to measure the difference between the original source and that of the decrypted image.

$$PSNR = 10\frac{255^2}{MSE}$$

Where MSE

$$MSE = \sum_{i=1}^{N} \sum_{j=1}^{M} \frac{[A\,(i,j) - B(i,j)]^2}{N \times M}$$

A lower PSNR value indicates a stronger image encryption method.

**Root Mean Squared Error (RMSE).** It is given as the square root of MSE. The root mean square error (RMSE) measures the amount of change per pixel due to the processing. The RMSE between a reference or original image, image1 -B (i, j) and the enhanced image, image2- A (i, j) is given by

$$RMSE = \sqrt{\frac{1}{M \times N} \sum_{i=0, j=0}^{M-1, N-1} [A(i,j) - B(i,j)]^2}$$

**Structural Similarity Index Measure (SSIM).** It is explained as

$$SSIM = \frac{(2\,a_A a_B + C_1).(2\,V_{AB} + C_2)}{(a_A^2 + a_B^2 + C_1).(a_A^2 + a_B^2 + C_2)}$$

where, $a_A$ and $a_B$ are the averages of the images A and B, respectively. Here, $V_A$ and $V_B$ represent the variance of A and B, respectively; whereas $V_{AB}$ is the covariance among A and B. Lastly, $C_1$ and $C_2$ are constants.

**Universal Quality Image Index (UQI).** It is a unique case of SSIM where $C_1 = C_2 = 0$

$$UQI = \frac{4 \cdot a_A a_B \cdot V_{AB}}{(V_A^2 + V_B^2) \cdot (a_A^2 + a_B^2)}$$

# 4    Architecture

The proposed architecture incorporates efficient security and data retrieval techniques. The architecture supports all kinds of 3D medical image formats for secure storage and faster retrieval.

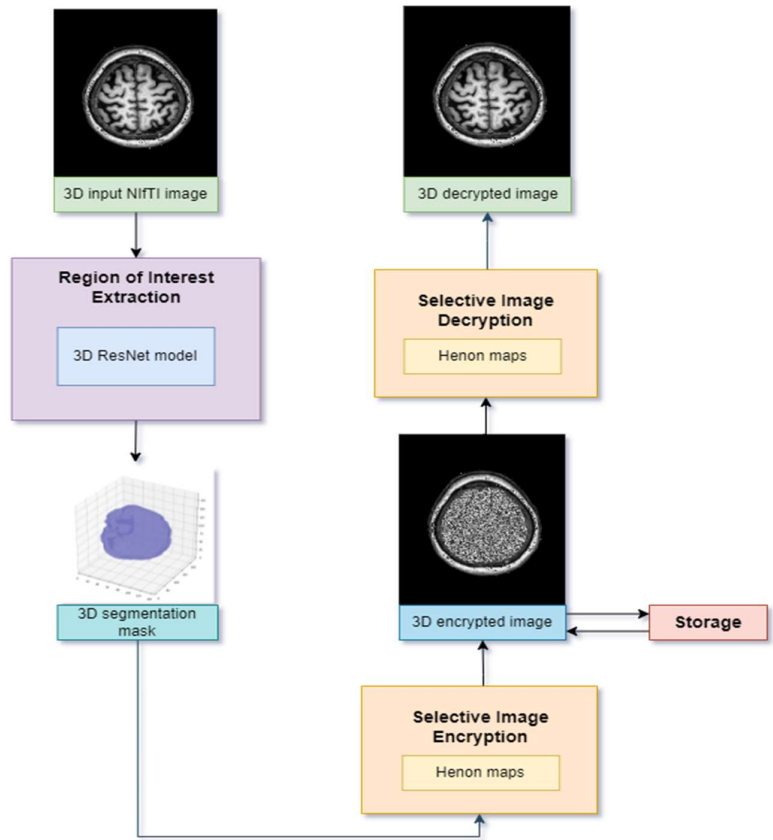

**Fig. 5.** Architecture Diagram

---

ALGORITHM: Selective Image Encryption Algorithm

---

```
inputVol <- 3D NIfTI Volume
spatialNormalizedVol <- spatialNormalization(inputVol)
intensityNormalizedVol <- intensityNormaliza-
tion(spatialNormalizedVol)
segmentationMask <- 3DResNet50(intensityNormalizedVol)
transformationMatrix <- generateChaoticMap(inputVol, encryp-
tionKey)
for each pixelPosition in inputVol do
  if segmentationMask[pixelPosition] == 1:
    inputVol[pixelPosition] <- transformationMa-
trix[pixelPosition] ^ inputVol[pixelPosition]
encVol <- inputVol
save encVol and segmentationMask in storage
transformationMatrix <- generateChaoticMap(encVol, encryp-
tionKey)
for each pixelPosition in encVol do
  if segmentationMask[pixelPosition] == 1:
    encVol[pixelPosition] <- transformationMa-
trix[pixelPosition] ^ encVol[pixelPosition]
decVol <- encVol
return decVol
end
```

It is necessary to normalize the inputted 3D NIfTI image in order for it to operate in the proposed ResNet architecture. A spatial normalization is employed to make the spatial distance in the image uniform. The median spacing between pairs of pixels is determined and used to redefine the distance between them. Intensity normalisation is also performed, which converts pixel intensity to a range of 0 to 1. To obtain the segmentation mask, the preprocessed image data matrix is fed into the ResNet architecture. This mask defines the region of interest that should be obscured. Encryption is performed by performing bitwise operation between image and a transformation matrix, obtained by generating a chaotic map with the help of a key and the input image matrix. The encrypted image is saved and may be accessed and decrypted at any time as per the user requests. Decryption employs the same technique as encryption, however the bitwise operation is performed between encrypted image matrix and the transformation matrix.

The whole process can be divided into 3 sections namely Region of Interest Extraction, Encryption & Storage, Decryption & Retrieval.

### 4.1 Region of Interest Extraction

The underlying segmentation architecture is a 3D ResNet-50 model which has been pre-trained using the Med3D framework. The benefit of the pre-trained Med3D models is the high segmentation accuracy arising from spatial and intensity normalisation performed by Med3D.

In ROI Segmentation the input medical image is organ, size or modality independent. The ROI segmentation was done using ResNet architectures. The following Resnets were tested in this scheme.

   a. Resnet 10
   b. Resnet 18
   c. Resnet 34
   d. Resnet 50

Resnet 50 outperformed other Resnets in whose results are discussed in the Results section. Resnet offers the model the capability to apply to various organs for generating the segmentation mask. The model could learn 3d spatial features from the medical images which provides an edge in better ROI detection.

By transfer learning, a patient specific mask is generated for individual use cases.

## 4.2 Encryption and Storage

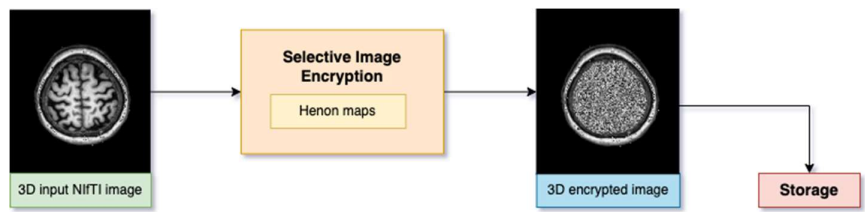

**Fig. 6.** Selective Encryption Unit

The masks generated are used to segment out the medically important parts of the scan image. To perform selective encryption, we adopt Henon maps. The "selective" feature in the selective encryption is facilitated by conditional checking for whether a region falls in the region of interest before performing encryption operation.

The Henon map algorithm uses a chaotic function and unique user defined key to generate a transformation matrix. The XOR of the transformation matrix with the original image on the region of interest is performed using a segmentation mask to produce the encrypted image. The encrypted image is stored for later retrieval.

### 4.3    Decryption and Retrieval

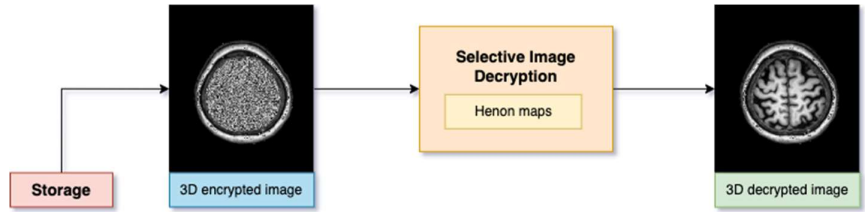

**Fig. 7.** Selective Decryption Unit

The encrypted images are fetched on demand by the user. The selective decryption takes the segmentation mask and selectively encrypted image as input. Once again, using Henon maps, conditionally, we decrypt the parts of the image that fall under the region of interest depicted in the segmentation mask.

The encrypted image is decrypted using the unique user defined key. The key is used to regenerate the transformation matrix with the chaotic function. The operation of XOR is performed between the transformation matrix and the region of interest in the encrypted image. By the XOR operation the original image is retrieved losslessly.

## 5    Methodology

To train the 3DResNet, 4 different models (ResNet-10, ResNet-18, ResNet-34 and ResNet-50) were trained. The images were resized to 512x512x38 during training. Batch size of 48 was used and the learning rate was adaptive. Cross Entropy Loss was used as the loss function and the models were trained over 110 epochs. Once the trained model is obtained, it is used to obtain segmentation masks of the input 3D medical image file.

The segmentation mask obtained from the previous step and the original image are passed into the encryption module. Here, the Henon map encryption is used with keys x=0.1 and y=0.1. This gives us the selectively encrypted 3D medical image file. A NumPy vectorised implementation is used to speed up this operation.

## 6    Results

Two distinct comparisons were conducted to determine the efficacy of the suggested strategy. RMSE, PSNR, UQI, and SSIM were the measures employed in this study. In the first comparison, the original image was compared with the encrypted image. In this case a high error value is desired as the encrypted image should not resemble the original. The below table depicts this comparison.

**Table 3.** Metrics between original and encrypted image.

| Metric | Value |
|--------|-------|
| RMSE | 44.998 |
| PSNR | 17.332 |
| UQI | 0.891 |
| SSIM | 0.746 |

The metrics measured between original image and the encrypted image are presented here. It can be observed from the RMSE value of 44.998 and PSNR value of 17.332 that the encryption introduces a lot of noise in the encryption process. The UQI value of 0.891 implies that the original image has been distorted to a decent degree with respect to loss of correlation, contrast distortion and illumination distortion. The SSIM value of 0.746 implies that there is a significant loss in structural similarity caused by the encryption process. The inference is that the encryption process leads to a significant degree of distortion in the encrypted image.

In the second comparison, the original image was compared against the decrypted image. A low error value is required in this scenario since the goal is to produce a decrypted image that closely resembles the original. This comparison is tabulated below.

**Table 4.** Metrics between original and decrypted image.

| Metric | Value |
|--------|-------|
| RMSE | 44.998 |
| PSNR | 17.332 |
| UQI | 0.891 |
| SSIM | 0.746 |

When the original image and final decrypted image are compared, it is observed that the encryption and decryption process together is lossless. Ideal values are obtained across RMSE, PSNR, UQI and SSIM (RMSE of 0.000, PSNR of infinity, UQI of 1.000 and SSIM of 1.000).

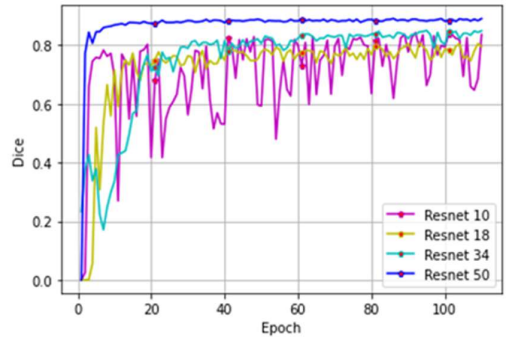

**Fig. 8.** Dice co-efficients for different ResNet architectures

When the dice coefficients were tracked over the span of epochs from 0 to 110, it is observed that in the end, the ResNet-50 is able to achieve the maximum segmentation accuracy (Dice coefficient) of 89.6%. In comparison, the other ResNet models are able to achieve a segmentation accuracy of 80% or a little above it. The primary benefit of the ResNet-50 model is that it is able to achieve its peak performance in close to 40 epochs. It is also to be noted that the ResNet-10 model achieves very unstable Dice coefficients throughout its training. This makes the ResNet-10 model unsuitable for use in our case.

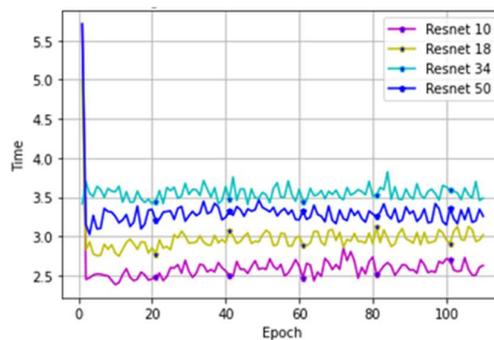

**Fig. 9.** Segmentation mask extraction times for different ResNet architectures

Next, observing the run times for generating segmentation masks, ResNet-34 takes the maximum prediction time, with the ResNet-50 following next. In ResNet-50 run time is 3.256 seconds, while in ResNet-34, run time is 3.504. The ResNet-18 takes 3.040 seconds and the ResNet-10 takes 2.658 seconds.

Epitomizing the result, the image has been adequately encrypted because there is a significant degree of distortion between original and encrypted image. The decryption obtained is also lossless. This scheme's ResNet architecture surpasses other architectures performance because it delivers excellent segmentation accuracy. Through the plots of percent gains in time for selective encryption vs full decryption, it was observed that selective encryption is 47% faster than full encryption during the decryption phase.

## 7 Future Scope

The future scope of this project would include adding lossless compression methods, transmission security and cloud storage which would further increase accessibility, security and storage resource optimisation. Introducing image processing and image enhancing techniques before ROI extraction would improve the ROI detection.

## 8 Conclusion

The current trend on medical image storage and security revolves around full encryption of medical images. The complete encryption of medical images pays off with its security by trading off on the computational resource utilisation. Though security is intact, the resource utilisation in the image retrieval by medical professionals for various purposes including diagnosis can be staggering high. The original image is scanned, encrypted and stored in a traditional way of medical image storage system. This scheme explores the Chaotic Henon Map based encryption algorithm selectively done on Medical Image Region of interest. The model built in this scheme uses Resnet architecture to segment the region of interest. The segmented ROI is selectively encrypted using a chaotic henon map encryption algorithm. The encryption function is generated using a unique user defined key. The encrypted ROI is stored for later retrieval. The computationally intensive work is done. Now every time a medical professional needs to retrieve the medical , the decryption algorithm is carried out using the unique user defined key. The algorithms proposed retrieves the medical image losslessly. By this algorithm the user is able to retrieve the medical image 47% faster saving time and computational resources. The underlying image array makes it possible for this architecture to handle all types of Medical Images (NIfTI, DICOM) making it versatile, faster and secure.